\documentclass[aps,pre,twocolumn,superscriptaddress]{revtex4}

\usepackage{graphicx}
\usepackage{amsmath}
\usepackage{amssymb}
\usepackage{color}
\usepackage{amscd}
\usepackage{bm}
\usepackage{enumerate}

\usepackage{mathrsfs}
\usepackage{calrsfs}
\usepackage{epsfig}
\usepackage{subfigure}
\usepackage{psfrag}

\def\beq{\begin{equation}}
\def\eeq{\end{equation}}
\def\bea{\begin{eqnarray}}
\def\eea{\end{eqnarray}}

\begin{document}
\title{Instabilities and Oscillations in Isotropic Active Gels}
\author{Shiladitya Banerjee}\thanks{sbanerje@syr.edu}\affiliation{ Physics Department, Syracuse University, Syracuse, NY 13244, USA}

\author{M. Cristina Marchetti}\thanks{mcm@physics.syr.edu}
\affiliation{Physics Department \& Syracuse Biomaterials Institute, Syracuse University, Syracuse, NY 13244, USA}

\date{\today}
\begin{abstract}

We present a generic formulation of the continuum elasticity of an isotropic  crosslinked active gel. The gel is described by a two-component model consisting of an elastic network coupled frictionally to a permeating fluid. Activity is induced by active crosslinkers that undergo an ATP-activated cycle and transmit forces to the network. The on/off dynamics of the active crosslinkers is described via rate equations for unbound and bound  motors. For large activity motors yield a contractile instability of the network. At smaller values of activity, the on/off motor dynamics provides an effective inertial drag on the network  that opposes elastic restoring forces, resulting in spontaneous oscillations. Our work provides a continuum formulation that unifies earlier microscopic models of oscillations in muscle sarcomers and a generic framework for the description of the large scale properties of isotropic active solids.
\end{abstract}
\maketitle

\section{Introduction} Much recent theoretical effort has focused on modeling the effect of
motor activity on the cell cytoskeleton. The cytoskeleton is a highly
heterogenous  polymer gel, mainly composed of filamentary actin
crosslinked by a myriad of globular proteins~\cite{AlbertsBook07}.
These include proteins that preserve the isotropic nature of the
network (e.g., filamin), proteins that induce bundle formation
(e.g., fascin or vilin), and molecular motor proteins, such as
kynesins and myosins, that are capable of transforming chemical
energy into mechanical work~\cite{HowardBook00}.  Motor proteins
hydrolyze adenosine-tri-phosphate (ATP) and convert it to
adenosine-di-phosphate(ADP) and inorganic phosphate(P). The free
energy released from this chemical reaction is used to generate
conformational changes of the motor proteins that yield mechanical
forces along cytoskeletal  filaments. The dynamics of the resulting
polymer network is controlled by active process on a range of time
scales, including the polymerization/deplymerization of the polar
filaments, the force-generation form crosslinking motor proteins,
and the load-dependent dynamics of these active crosslinkers.

Theoretical work has modeled the cytoskeleton via generic continuum
hydrodynamics as an \emph{active liquid}, where the effect of
activity is incorporated via suitable modification of the
hydrodynamic equations of equilibrium liquid
crystals~\cite{Kruse2004,Kruse2005,JKPJ-PhysRep2007}. The continuum
theory has led to several predictions, including the onset of
spontaneous deformation and flow in  active
films~\cite{Voituriez06,GiomiMarchettiLiverpool2008}, the formation
of spiral and aster patterns reminiscent of those observed in
in-vitro extracts of cytoskeletal filaments and motor
proteins\cite{Kruse2004,Kruse2005,Ndlec1997,Surrey2001,Aranson2005,Aranson2006,Liverpool2003,Ahmadi2005},
and activity-induced thinning and thickening in sheared active
suspensions~\cite{Hatwalne04,TBLMCM06,GiomiMarchettiLiverpool2010,Cates2008}.
Viscoelasticity has also been incorporated in the continuum theory
using the Maxwell model that modifies the response of the liquid by
introducing a characteristic time scale controlling the crossover
from fluid behavior at long times to elastic behavior at short times
~\cite{JKPJ-PhysRep2007}. Given, however, that the active liquid
viscoelastic model cannot support elastic stresses at long times,
its direct relevance for the understanding of the crawling dynamics
of the lamellipodium and of active contractions in living cells
remains to be established. In addition, the active liquid model is
inadequate to describe cross-linked contractile systems, such as
stress fibers (cross-linked bundles of actin filaments and myosin
minifilaments that play a crucial role in controlling the ability of
non-muscle animal cells to generate and resist
forces)~\cite{Pellegrin2007} or muscle sarcomeres that often exhibit
spontaneous oscillations~\cite{Anazawa1992}. Such oscillations
require long-wavelength elastic restoring
forces~\cite{JulicherProst1997,GuntherKruse2007} not accounted for
in an active (even viscoelastic) liquid. This suggests that the
long-wavelength properties of stress fibers or sarcomers may be
better described as those of an active elastic medium or
\emph{active solid}. Polarity is generally expected to also play an
important role in these systems indicating that a suitable continuum
model maybe that of an active polar elastomer gel.

Passive polymer gels are often classified on the basis of the nature
of the crosslinking forces~\cite{Rubinstein2003}. Chemical gels have
strong cross-links bound by covalent bonds. These crosslinks have an
essentially infinite lifetime on all experimentally relevant time
scales and the gel behaves elastically at long times, with a finite
shear modulus. At short times, however, dissipation induced by
internal frictional processes can result in  ``liquid-like"
response, with the loss (viscous) component of the elastic moduli
exceeding the storage (elastic) component.  In physical gels, in
contrast, the crosslinks are held together by weaker  interactions
(e.g., dipolar or ionic) and have finite lifetimes, ranging from
minutes to a fraction of a second. This yields a broad spectrum of
behavior, from strong physical gels, that are similar to chemical
gels, to weak physical gels, with reversible links formed by
temporary associations between chains. The latter are liquid at long
time and exhibit elasticity on short time scales.

Similarly, active polymer gels also may or may not exhibit low
frequency elasticity, depending on the nature of the crosslinkers.
Cross-linked reconstituted actin networks exhibit some of the
properties of strong physical gels and display large active
stiffening driven by molecular motors~\cite{Mizuno2007}.
MacKintosh and Levine~\cite{MacKintoshLevine2008,LevineMacKintosh2009} and
Liverpool et al.~\cite{LMJP-EPL2009}  showed that elastic networks
with contractile forces induced by myosin II motors, described as
static force dipoles, can account for both the large scale
contractility and stiffening observed in experiments. In a recent
paper G\"unther and Kruse~\cite{GuntherKruse2007}  also demonstrated
that a continuum theory obtained by coarse graining a specific
microscopic model of coupled sarcomeres does yield oscillatory
states, as observed ubiquitously in these systems, provided the
load-dependent on/off dynamics of motor proteins is included in the
hydrodynamic model.  Motor proteins are also directly involved in
controlling mechanical oscillations and instabilities in cilia and
flagella~\cite{Brokaw1975,Camalet1999,Camalet2000} and in the
mitotic spindle during cell division~\cite{Grill2005}. In all these
cases the elastic nature of the network at low frequency is crucial
to provide the restoring forces need to support oscillatory
behavior, i.e., these systems are best modeled as active solids,
rather than active liquids.

In this paper we formulate a generic continuum theory of
\emph{isotropic} cross-linked active gels that incorporates the
on/off dynamics of crosslinking motor proteins. Following MacKintosh
and Levine~\cite{MacKintoshLevine2008,LevineMacKintosh2009}, we
model the gel as a two-component system composed of an elastic
network coupled frictionally to a permeating fluid. The details of
the model are given in section II. The active forces arising from
motor proteins are incorporated phenomenologically through an active
contribution to the stress tensor of the elastic network and are
controlled by the load-dependent on/off dynamics of the motors. In
section III we examine the hydrodynamic modes of the active gel and
show that, as stated in Ref.~\cite{GuntherKruse2007}, a large
activity can change the sign of the effective compressional modulus,
yielding a contractile instability. Spontaneous oscillations are
obtained in the regime of weak activity where the compressional
modulus is softened by bound motors, but remains positive. In
section IV we consider the case of an overdamped gel relevant to
muscle fibers and show that it can exhibit propagating waves and
oscillatory instabilities as parameters are varied. A phase diagram
summarizing the behavior is given in Fig. 2. In Section V we
describe the macroscopic homogeneous response of the active medium
as probed in creep experiments and by macroscopic rheology
measurements. The two-component gel model exhibits viscous response
on short time scales and elastic response at long times
~\cite{Levine2001} even in the absence of activity, when the time
scale controlling the crossover between these two responses is set
by  the ratio of the viscosity and the compressional modulus of the
network. Activity renormalizes the time scale controlling this
crossover. Finally, we conclude with a brief discussion.

\section{Hydrodynamics of Isotropic Active Gels}
Hydrodynamics is a systematic method to study the behavior of
extended systems on long times and length scales by focusing on the
dynamics of conserved and broken symmetry fields. Here we use a
phenomenological symmetry-based approach to formulate a continuum
hydrodynamic description of a cross-linked gel (e.g., a network of
actin filaments crosslinked by filamins or other "passive" linkers)
under the influence of active forces exerted by clusters of
crosslinking motor proteins (e.g., myosin II minifilaments). We
consider a three-dimensional isotropic polymer gel of mesh size
$\xi$, viscously coupled to an incompressible permeating Newtonian
fluid~\cite{Levine2001}. This two component model has been used
previously to determine viscoelastic response of a filamentous
isotropic network in
solution~\cite{Levine2001,LevineLubensky2000,MacKintoshLevine2008,LevineMacKintosh2009},
and more recently to discuss mechanical response of a coupled
network-solvent system when probed by an active
agent~\cite{Head2010}. At length scales larger than $\xi$ the
deformations of the polymer network can be described by isotropic
elasticity in terms of a continuum displacement field, ${\bf u}({\bf
r},t)$ and an elastic free energy given by
\begin{equation}
F_e=\frac{1}{2}\int_{{\bf r}}\left( \lambda u_{ii}^2 + 2\mu u_{ij}u_{ij} \right)\;,
\label{Fe}
\end{equation}
with $\lambda$ and $\mu$ the usual bulk and shear Lam\'e coefficients and $u_{ij}=\frac{1}{2}(\partial_i u_j +\partial_j u_i)$  the strain tensor.
The permeating viscous fluid is characterized by a velocity field ${\bf v}({\bf r},t)$ and the coupling between the network and the fluid is controlled by a friction per unit volume, $\Gamma$.  The equation of motion for the displacement field can be written as
\begin{equation}
\rho \ddot{{\bf u}}=-\Gamma (\dot{{\bf u}}-{\bf v})+\bm\nabla\cdot\bm\sigma \;,\label{udot}
\end{equation}
where $\rho$ is the mass density of the network and $\bm\sigma$ is the stress tensor of the gel.
The permeating fluid is described by the  Navier-Stokes equation,
\begin{equation}
\rho_f\dot{{\bf v}} - \eta\nabla^2{\bf v} + \nabla P = \Gamma\left(\dot{{\bf u}}-{\bf v}\right)\label{vdot}
\end{equation}
where $\rho_f$ is the mass density of the fluid, $\eta$ the fluid shear viscosity,
and $P$ is the pressure. We have assumed a low Reynolds number regime for the fluid and omitted the convective term from the Navier-Stokes equation. It is also assumed that motor proteins do not exert any direct forces on the permeating fluid.
As discussed elsewhere~\cite{Levine2001}, the friction $\Gamma$ between the elastic network and the permeating fluid can be estimated by considering a polymer strand of length $\xi$ moving relative to the background fluid at a velocity $v$.  By equating the  viscous force density $\sim\eta v/\xi^2$ on the strand  to the viscous friction $\sim\Gamma v$ due to the permeating fluid, one obtains an  estimate of the friction as $\Gamma \sim \eta/\xi^2$. The frictional drag per unit volume $\Gamma$ is then determined by the force density required to drive a fluid of viscosity $\eta$ through network pores of characteristic cross section $\xi^2$.

The stress tensor of the gel can be written as the sum of elastic, dissipative and active parts,
\beq
\bm\sigma=\bm\sigma^e +\bm\sigma^d + \bm\sigma^a\;.
\eeq
The elastic contribution  is given by $\sigma_{ij}^e=\frac{\delta F}{\delta u_{ij}}$, with
\begin{equation}
\sigma_{ij}^e=\left(\lambda+\frac{2 \mu}{3}\right)\delta_{ij}\bm\nabla\cdot{\bf u} + 2\mu\left(u_{ij} - \frac{1}{3} \delta_{ij}\bm \nabla\cdot{\bf u}\right)\;.
\end{equation}
The dissipative component $\bm\sigma^d$ is given by
\begin{equation}
\sigma_{ij}^d=\eta_b \delta_{ij} \bm\nabla\cdot\dot{{\bf u}} + 2 \eta_s \left( \dot{u}_{ij} - \frac{1}{3} \delta_{ij} \bm\nabla\cdot\dot{{\bf u}}\right)\;,
\end{equation}
where $\eta_b$ and $\eta_s$ are bulk and shear viscosities arising from internal friction in the gel.
Changes in the density $\rho$ of the network are slaved to changes in volume, thus $\delta\rho=-\rho_0 \bm\nabla\cdot{\bf u}$, with $\rho_0$ the mean mass density of the elastic network.  In addition, we neglect here for simplicity energy fluctuations and assume that the fluid surrounding the network serves as a heat bath and maintains the temperature constant. This approximation is not adequate to describe real muscle fibers that heat upon contraction.

The active contribution, $\bm\sigma^a$, to the stress tensor  arises from the forces exerted by motor proteins bound to the filaments. We assume a total concentration $c=c_b+c_u$ of motor proteins in the gel, with $c_b$ and $c_u$ the concentrations of bound and unbound motors, respectively. In an isotropic network the active contribution to the stress tensor can generically be written as ~\cite{Kruse2005},
\begin{equation}
\sigma_{ij}^a=\delta_{ij}~\zeta(\rho, c_b)~\Delta \mu \;, \label{actsigma}
\end{equation}
where $\Delta \mu$ is the change in chemical potential due to the hydrolysis of ATP and $\zeta(\rho, c_b)$ is a scalar function with dimensions of number density describing the stress per unit change in chemical potential due to the action of active crosslinkers.

To complete the hydrodynamic description we need equations describing the dynamics of bound and unbound motors.  We assume unbound motors diffuse in the permeating fluid, while bound motors  are convected with the polymer network. Their dynamics is controlled by prescribed binding and unbinding rates, $k_{b}$ and $k_{u}$ according to first-order reaction kinetics. The resulting equations are
\begin{equation}
\partial_t{c_b}+ \bm\nabla\cdot\left( c_b \dot{{\bf u}}\right)= -k_{u}c_b + k_{b}c_u \label{cb}\;,
\end{equation}
\begin{equation}
\partial_t{c_u} = D\nabla^2 c_u + k_{u}c_b - k_{b}c_u \label{rhom}\;,
\end{equation}
where $D$ is the diffusion coefficient for free motors. The rates
$k_{b}$ and $k_{u}$ depend of course on the specific type of motor
protein considered. Each motor protein undergoes a conformational
transformation during a cycle fueled by a chemical reaction,
generally the hydrolysis of ATP~\cite{HowardBook00}. The total cycle
duration is determined by the sum of the time $\tau_{on}$  that the
protein spends attached to the filament, doing its working stroke,
and the time $\tau_{off}$ that it spends detached from the filament,
making its recovery stroke. Motor proteins are generally
characterized by the value of the duty ratio,
$r=\tau_{on}/(\tau_{on}+\tau_{off})$. Myosins II with $r\sim
0.05$~\cite{HowardBook00} spend most of their time unbound, while
two-headed kinesins have values of $r$ close to unity and are
classified as highly processive motors that remain attached to the
filament for most of the duration of the cycle. The binding and
unbinding rates are then estimated as $k_{u}\sim 1/\tau_{on}$ and
$k_{b}\sim 1/\tau_{off}$. For individual myosins II, $\tau_{on}\sim
2ms$ and $\tau_{off}\sim 40ms$~\cite{HowardBook00}, corresponding to
$k_{b}\ll k_{u}$. During the working stroke and in the absence of
external load, the protein moves along the filament at a speed
$v_0\sim\Delta\mu$. The time $\tau_{on}\sim 1/v_0\sim1/\Delta\mu$
depends on motor activity, while $\tau_{off}$ is essentially
independent of $\Delta\mu$. For myosins II this gives
$k_u\sim\Delta\mu\gg k_b$.

Finally, we assume that the four-component active gel described by
the set of coupled equations \eqref{udot}, \eqref{vdot}, \eqref{cb}
and \eqref{rhom} is incompressible. This requires
\begin{equation}
\bm\nabla\cdot[\left(1-\phi_p\right){\bf v} + \phi_p\dot{\bf u}]=0\;, \label{continuity}
\end{equation}
where $\phi_p$ denotes the combined volume fraction of the polymer
network  with bound motors. We assume that the volume fraction of
the network is very small, i.e. $\phi_p << 1$. In this case
Eq.~(\ref{continuity}) reduces to the condition of incompressibility
of the ambient fluid, $\bm\nabla\cdot{\bf v}\simeq0$.

In the homogeneous steady state the network and fluid densities have constant values $\rho_0$ and $\rho_f$, respectively. The relative  concentrations of bound and free motors are controlled  by the binding/undinding rates and are given by
\begin{subequations}
\begin{gather}
c_{b0}=\frac{k_b}{k_b+k_u}c_{m0}\;,\label{cb0}\\
c_{u0}=\frac{k_u}{k_b+k_u}c_{m0}\;,\label{cu0}
\end{gather}
\end{subequations}
with $c_{m0}$ the total steady state concentration of motor
proteins. In the following we are mainly interested in
non-processive motors like myosins II that are mostly unbound on
average, with $c_{b0}<<c_{u0}$. In this case we neglect the dynamics
of free motors that essentially provide a "motor reservoir" and
assume that $c_u\sim c_{u0}$ in \eqref{cb}. In addition, we expand
$\zeta(\rho,c_b)$  to linear order in fluctuations of  the network
density and motor concentration from their equilibrium values,
$\delta\rho=\rho-\rho_0$ and $\delta c_b=c_b-c_{b0}$, as
\begin{equation}
\zeta(\rho, c_b)=\zeta_0 + \zeta_1 \frac{\delta\rho}{\rho_0} +
\zeta_2\frac{\delta c_b}{c_{b0}}\;. \label{zetaexp}
\end{equation}
The microscopic parameter $\zeta_0$ is related to a stall force, but will not play a role in the following. The parameter $\zeta_1$ arises from spatial variations in the motor density. Both $\zeta_1$ and $\zeta_2$ are expected to be positive for contractile systems.

\section{Hydrodynamic Modes and Linear Stability of Homogeneous Stationary State}
In this section we consider the linear stability of the homogeneous stationary state, with ${\bf u}={\bf v}=0$, $\rho=\rho_0$ and $c_b=c_{b0}$ by examining the hydrodynamic modes of the incompressible gel.
 The fluid density $\rho_f$ is fixed due to the condition of incompressibility.
Using \eqref{zetaexp} for the active parameter $\zeta$, the linearized hydrodynamic equations are  given by
\begin{subequations}
\begin{gather}
\rho_0 \ddot{{\bf u}}-\mu \nabla^2 {\bf u} - (\lambda + \mu -\zeta_1\Delta\mu) \bm\nabla (\bm\nabla\cdot{\bf u}) =
\Gamma\left({\bf v}-\dot{{\bf u}}\right)\notag\\
+  \eta_s \nabla^2 \dot{{\bf u}} + \left(\eta_b + \frac{\eta_s}{3}\right) \bm\nabla (\bm\nabla\cdot\dot{{\bf u}}) + \zeta_2\Delta\mu \bm\nabla\phi\;, \label{ud}\\
\rho_f\dot{{\bf v}} - \eta\nabla^2{\bf v} + \bm\nabla P = \Gamma\left(\dot{{\bf u}}-{\bf v}\right)\;, \label{eqv}\\
\dot{\phi}+ \bm\nabla\cdot\dot{{\bf u}}=-k_{u}\phi \;,\label{delrhob}
%
\end{gather}
\end{subequations}
with $\phi=\delta c_b/c_{b0}$ and  the  condition $\bm\nabla\cdot{\bf v}=0$.

We now discuss the hydrodynamic modes of the three-component system
described by Eqs.~(\ref{ud}-\ref{delrhob}) obtained by neglecting
fluctuations in free motors. We expand the fluctuations $\delta
y_\alpha=\left({\bf u},{\bf v},\phi\right)$ in Fourier components
according to
\beq
\delta y_\alpha ({\bf r},t)=\int_{\bf q}e^{-i{\bf q}\cdot{\bf r}}~\delta \tilde{y}_\alpha({\bf q},t)
\eeq
and look for solutions with time dependence of the form
$\delta\tilde{y}_\alpha({\bf q},t)\sim e^{-i\omega
t}\delta\tilde{y}_\alpha({\bf q})$. We also write ${\tilde{\bf u}}$
into its components transverse and longitudinal to ${\bf q}$ by
letting $\tilde{\bf u}=\hat{\bf q}u_L+\tilde{\bf u}_T$, with
$\hat{\bf q}={\bf q}/q$ and $\hat{\bf q}\cdot\tilde{\bf u}_T=0$. Due
to incompressibility of the background fluid, $\tilde{\bf v}$ does
not have any longitudinal component, and incompressibility allows us
to eliminate the pressure $P$ from (\ref{eqv}). In Fourier space,
dropping for simplicity of notation the tilde on the Fourier
components of the fluctuations, the equations for the longitudinal
fluctuations are given by
\begin{subequations}
\begin{gather}
\left[-\rho_0 \omega^2 + (B-\zeta_1\Delta\mu) q^2 -\imath\omega(\Gamma+ \eta_L  q^2)\right]u_L= - \imath q\zeta_2\Delta\mu\phi \;, \label{uL}\\[-5pt]
(-\imath \omega +k_{u})\phi= \omega  qu_L \;, \label{rhob}
\end{gather}
\end{subequations}
where we  have defined the longitudinal modulus of the gel as
$B=\lambda+2\mu$ and a longitudinal viscosity of the network as
$\eta_L=\eta_b+(4/3)\eta_s$. The longitudinal part of the
displacement couples to motor density, but not to the velocity of
the permeating fluid in the incompressible limit considered here.
Fluctuations in the longitudinal displacement are slaved to
fluctuations in the network density, with $\delta\rho=\rho_0 iq
u_L$. The longitudinal equations (\ref{uL}) and (\ref{rhob}) can
then also be rewritten as coupled equations for fluctuations in the
network and bound motor densities,
\begin{subequations}
\begin{gather}
\left[-\rho_0 \omega^2 + (B-\zeta_1\Delta\mu) q^2 -\imath\omega(\Gamma+ \eta_L  q^2)\right]\frac{\delta\rho}{\rho_0}= \zeta_2\Delta\mu q^2 \phi\;, \label{deltarho}\\[-5pt]
(-\imath \omega +k_{u})\phi =- \imath \omega \frac{\delta\rho}{\rho_0} \;, \label{rhobrho}
\end{gather}
\end{subequations}
%
%
Finally, the equations for the transverse components are given by
\begin{equation}
\left[-\rho_0 \omega^2 -\imath\omega(\Gamma+\eta_s q^2)+\mu q^2\right]{\bf u_T}= \Gamma {{\bf v}} \label{uT}\;,
\end{equation}
\begin{equation}
\left(-\imath \omega\rho_f + \Gamma+\eta q^2\right){{\bf v}}=-\imath \omega \Gamma {\bf u_T} \;.\label{vT}
\end{equation}
and are decoupled from the  equations for the longitudinal modes. We therefore proceed to analyze the two groups separately.

\subsection{Longitudinal  Modes}
In the incompressible limit considered here, the only role of the permeating fluid is to provide the frictional damping $\Gamma$. The longitudinal deformations of the polymer network do, however,  couple to fluctuations in the bound motor density. It is instructive to first review the behavior of a passive gel, as obtained by letting $\Delta \mu=0$ in Eq.(\ref{uL}).

\subsubsection{Passive gel.}
\begin{figure}[h]
\begin{center}
\includegraphics[scale=0.85]{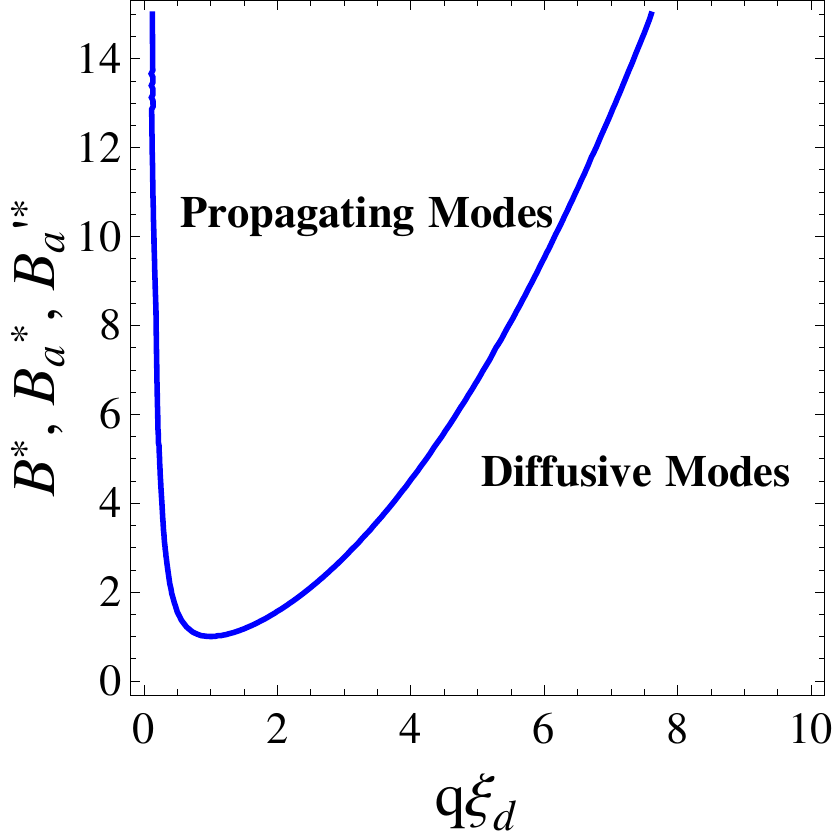}
\end{center}
\caption{The critical line  $B^*(\tilde{q})$ given in
Eq.~\eqref{Estar} for $\epsilon=1$ separating the region of
parameters where density fluctuations in a passive incompressible
gel relax  diffusively, from the region where the system supports
propagating density waves. In the chosen dimensionelss units, the
same line also describes the boundary $B_a^*(\tilde{q})$ obtained
for the case $k_u\rightarrow\infty$ and boundary $B_a^{\prime
*}(\tilde{q})$ obtained for the case $k_u=0$.}

\label{fig:PDpassive}
\end{figure}

\hspace{0.05in}
In the absence of motor proteins,  longitudinal fluctuations in an incompressible gel are  controlled by a single equation, given by
\begin{equation}
\left\{-\rho_0 \omega^2 +B q^2 -\imath\omega\left[\Gamma+\eta_L
q^2\right]\right\}u_L= 0\;. \label{deltarho-passive}
\end{equation}
We stress that this equation also describes the behavior of fluctuations sin the network density, as
$\delta\rho=\imath q\rho_o u_L$.  The hydrodynamic modes are the roots of the quadratic polynomial in curly brackets in Eq.~(\ref{deltarho-passive}) and are given by
\begin{equation}
\omega=-\frac{\imath}{2\rho_0}\left[\Gamma+\eta_L
q^2\pm\sqrt{(\Gamma+\eta_L q^2)^2-4\rho_0 B q^2}\right]\;.
\label{modesL-passive}
\end{equation}
The behavior is controlled by the interplay of two length scales, $\xi_d=\sqrt{\eta_L/\Gamma}$, the length scale over which intrinsic viscous dissipation within the network is comparable to dissipation due to friction with
the permeating fluid, and
$\ell_\Gamma=2\sqrt{\rho_0 B}/\Gamma$ controlling the ratio of
elastic restoring forces in the network to viscous drag from the
permeating fluid. The length scale $\ell_\Gamma$ has been introduced
before by Mackintosh and Levine~\cite{MacKintoshLevine2008,LevineMacKintosh2009}. At small
wavevector ($q\ll \ell_\Gamma^{-1}$) the dispersion relations are
always imaginary, corresponding to relaxational or diffusive modes,
and take the form
\begin{eqnarray}
\label{modesL0}
&&\omega_{L,\Gamma}^0=-\imath\left[\frac{\Gamma}{\rho_0}+\left(\frac{\eta_L}{\rho_0}-\frac{B}{\Gamma}\right)q^2\right]+O(q^4)\;,\\
&&\omega_L^0=-\imath\frac{B}{\Gamma}q^2+O(q^4)\;,
\end{eqnarray}
where the superscript $0$ is used to denote the passive gel limit. The mode $\omega_{L,\Gamma}^0$ is non-hydrodynamic and describes the relative motion of the polymer network and the permeating fluid. The mode $\omega_L^0$ describes the diffusive relaxation of network density fluctuations. In the two-fluid incompressible gel model considered here there are no propagating longitudinal sound waves \cite{Levine2001} and  the network density $\delta\rho$ relaxes diffusively, while the solvent density $\rho_f$ remains fixed. The limit $q\ll \ell_\Gamma^{-1}$ holds if $\ell_\Gamma<\xi_d$. On the other hand, when $\ell_\Gamma>\xi_d$, the modes are relaxational as given in Eqs.~\eqref{modesL0} for $q\ll \ell_\Gamma^{-1}$, but there is an intermediate regime of $\ell_\Gamma^{-1}<q<\xi_d^{-1}$ where
 the gel can support  propagating sound-like density waves.
Propagating density waves exist if  the argument of the square root
on the right hand side of Eq.~\eqref{modesL-passive} is positive,
i.e., $4\rho_0 B q^2>[\Gamma+\eta_Lq^2]^2$.  It is convenient to
scale lengths by $\xi_d$, with $\tilde{q}=q\xi_d$. The
condition for the existence of propagating waves can then be written
as
\begin{equation}
B^*\geq\frac{(1+\epsilon\tilde{q}^2)^2}{4\tilde{q}^2}\;,
\label{Estar}
\end{equation}
where $B^*=\frac{B\rho_0}{\Gamma^2\xi_d^2}$ and
$\epsilon=\eta_L/\eta$. The propagating waves are controlled by  the
interplay of inertia and elasticity and decay on time scales of the
order of the relaxation time  $\tau_\Gamma=\rho_0/\Gamma$, which is
set by the frictional damping from the solvent. The equality sign in
Eq.~\eqref{Estar} defines the critical line shown in
Fig.~\ref{fig:PDpassive}  separating the region of diffusive density
relaxation from the region where the system supports   propagating
sound-like waves.
No propagating waves exist for $B<\Gamma^2\xi_d^2/\rho_0$,
corresponding to the minimum of the curve in
Fig.~\ref{fig:PDpassive}. We stress that the modes are always
diffusive at the longest wavelengths, when $q\rightarrow 0$. These
finite wavevector sound-like waves persist down to very small
wavevector in the limit  of vanishing friction $\Gamma$ with the
surrounding fluid. This is seen by setting $\Gamma=0$  first,
followed by the small wavevector approximation. The dispersion
relations then take the form
\begin{equation}
\omega_{\pm}^0=\pm q\sqrt{\frac{B}{\rho_0}}-\imath
q^2\frac{\eta_L}{2\rho_0}\;. \label{omega-pm0}
\end{equation}
These are indeed sound waves propagating at the longitudinal sound
speed $\sim\sqrt{B/\rho_0}$.

\subsubsection{Neglecting bound motor fluctuations ($k_u\rightarrow\infty$).}\hspace{0.05in}
We now proceed to incorporate the effect of motor proteins. We first
consider the case of stationary bound motors. This can be obtained
in two ways, either by letting $k_u\rightarrow\infty$, which
corresponds to neglecting bound motor fluctuations, or by letting
$k_u=0$, which corresponds to neglecting the motor on/off dynamics.
In both limits motor activity can yield a contractile instability of
the system, but no spontaneous oscillations, as pointed out in
Ref.~\cite{GuntherKruse2007}.

When $k_u\rightarrow\infty$, then $\phi=0$ and the concentration of bound motors is constant, $c_b=c_{b0}$.  We then obtain a single decoupled equation for fluctuations in the longitudinal displacement (or equivalent, in the network density $\delta\rho$) of the form
\begin{equation}
\left\{-\rho_0 \omega^2 +(B-\zeta_1\Delta\mu) q^2
-\imath\omega\left[\Gamma+\eta_L q^2\right]\right\}u_L\;.
\label{deltarho1}
\end{equation}
In this  limit the only effect of motor activity is a contractile reduction of the compressional modulus, which is given by
\begin{equation}
B_{a}=B-\zeta_1\Delta\mu\;.
\end{equation}
  The hydrodynamic modes are identical to those described in the previous subsection, with the replacement $B\rightarrow B_{a}$. If $B_a<0$ the imaginary part of the mode $\omega_L^0$ changes sign, signaling a contractile instability of the system driven by motor activity.
When $B_a>0$, the modes can be real at finite wavevector,
corresponding to propagating waves. The condition for the existence
of propagating waves is precisely as given in Eq.~\eqref{Estar} for
the passive gel, with the replacement $B\rightarrow B_a$. A plot of
$B^*_a=B_a\rho_0/(\Gamma^2\xi_d^2)$ as a function of $\tilde{q}$ is
that identical to that shown in Fig.~\ref{fig:PDpassive} for the
passive case.
 We stress that the existence of these propagating density waves is {\em not} a consequence of activity. There is in fact a maximum value of activity, given by $\zeta_1\Delta\mu_c=B-\Gamma\xi_d^2/\tau_\Gamma$, and corresponding to the minimum of the curve plotted in Fig.~\ref{fig:PDpassive} above which there are no propagating modes. In addition, since $\ell_\Gamma\sim\sqrt{B_a}$ {\em decreases} with increasing activity $\Delta\mu$, the range of wavevectors where propagating waves exist for a fixed $B_a^*$ decreases with increasing activity and is given by
$\Delta\tilde{q}=2[B_a^*(B_a^*-1)]^{1/4}$.

\subsubsection{Neglecting bound motor dynamics ($k_u=0$).}\hspace{0.05in}
In this case bound motors remain bound at all times and bound motor fluctuations are slaved to network density fluctuations, with $\phi=\delta\rho/\rho_0$. The relaxation of longitudinal fluctuations is described by
\begin{equation}
\left\{-\rho_0 \omega^2 +\left[B-(\zeta_1+\zeta_2)\Delta\mu\right]
q^2 -\imath\omega\left[\Gamma+\eta_L q^2\right]\right\}u_L= 0\;.
\label{deltarho1}
\end{equation}
and the only effect of static bound motors is a further downward renormalization of the elastic modulus, which is now given by
\begin{equation}
B_{a}^\prime=B-(\zeta_1+\zeta_2)\Delta\mu\;.
\end{equation}
The modes are again formally identical to those obtained for the
passive gel, but with $B\rightarrow B_a^\prime$. The gel exhibits a
contractile instability for $B_a^\prime<0$ and finite wavevector
progating density waves for $B_a^\prime>0$. Note that the limit
where all motors are bound  can be obtained for instance after full
hydrolysis of ATP to ADP. Myosin has a high affinity to actin, hence
in a pure ADP environment it will act as a ``permanent" bound
crosslinker~\cite{Humphrey2002}. In this case, however, there will
also be no reduction of the elastic modulus due to activity, hence
no contractile instability. In fact muscles become rigid as ATP runs
out, which is one of the causes of \emph{rigor mortis}.

\subsubsection{Including bound motors dynamics (finite $k_u$).}\hspace{0.05in}
 We now incorporate the dynamics of the bound motors  and consider the hydrodynamic modes of the the two coupled equations (\ref{deltarho}) and (\ref{rhobrho}). These yield a cubic   eigenvalue equation, given by
\begin{eqnarray}
\lefteqn{\!\!\!\!\!\!\imath \rho_0 \omega^3 - \omega^2(\Gamma + k_{u}\rho_0+\eta_Lq^2) -\imath \omega \Big\{k_{u}(\Gamma+\eta_L q^2)}\nonumber\\
 &&+ \left[B-\left(\zeta_1+\zeta_2\right)\Delta\mu\right]q^2\Big\} + k_{u}B_a q^2=0\;. \label{omega}
 \end{eqnarray}
 The behavior is now controlled by the competition of two time scale, the network relaxation time $\tau_\Gamma=\rho_0/\Gamma$ and the time scale $\tau_{on}=k_u^{-1}$ characterizing the motors on/off dynamics.
 Solving perturbatively for small wave numbers q, the three modes  are given by
 \begin{equation}
 \label{omL}
 \omega_L=-\imath \frac{B_a}{\Gamma} q^2 + O(q^4)\;,
 \end{equation}
 \begin{equation}
 \label{omb}
 \omega_b=-\imath k_{u} + \imath q^2 \frac{\zeta_2\Delta\mu}{\Gamma -k_{u}\rho_0} + O(q^4)\;,
 \end{equation}
 %
 %
 %
 \begin{equation}
 \label{omLG}
 \omega_{L,\Gamma}=-\imath \frac{\Gamma}{\rho_0} +\imath q^2 \left(\frac{B_a}{\Gamma} - \frac{\zeta_2\Delta\mu}{\Gamma-k_{u}\rho_0}-\frac{\eta_L}{\rho_0}\right) + O(q^3)
 \end{equation}
 The mode $\omega_L$ describing the relative mass diffusion of network and solvent in the gel is unchanged at small wavevector. Again, it  changes sign when  $\Delta\mu > B/\zeta_1$, corresponding to a contractile instability of the gel that occurs when the active stresses exceed the elastic restoring forces from the passive elements of the polymer network.
 The other two modes are non-hydrodynamic and always stable at long wavelengths. The mode with relaxation rate $\omega_b$ describes the decay of fluctuations in the density of bound motors.  The mode with relaxation rate $\omega_{L,\Gamma}$ describes the damping of the network due to its motion with respect to the permeating fluid. Even when the on/off dynamics of the bound motors is taken into account,  no spontaneous oscillations are generated by motor activity \emph{in the long wavelength limit}. Oscillatory solutions do, however, occur at finite wavevector, as described below. We note that, although the modes always remain stable, the coupling to motor activity can yield a change in sign of the ${\cal O}(q^2)$ damping in $\omega_b$ and $\omega_{L,\Gamma}$. This effective "negative viscosity" due to motors occurs when the time scale of the motor on/off dynamics is fast compared to the  frictional relaxation of the network, i.e., for    $\tau_\Gamma>\tau_{on}$.   This "negative friction" effect of motors will become important below and was also discussed in Prost et al~\cite{JulicherProst1995}.

As in the passive case, the dispersion relations of the hydrodynamic
modes of our model viscoelastic gel depend on the order in which the
limits $\Gamma\rightarrow 0$ and $q\rightarrow 0$ are taken.  Above
we considered the small $q$ limit for fixed $\Gamma$. If in contrast
we take $\Gamma\rightarrow 0$ first, followed by  $q\rightarrow 0$
we obtain propagating modes (for $B_a>0$). The mode $\omega_b$
describing relaxation of bound motor fluctuations is qualitatively
unchanged and takes the form
 \begin{equation}
 \omega_{b}= -\imath k_{u} - \imath q^2 \frac{\zeta_2\Delta\mu }{k_{u}\rho_0} + O(q^4)\;.
 \end{equation}
The two modes $\omega_L$ and $\omega_{L,\Gamma}$ describing the
dynamics of network density fluctuations are replaced by two
propagating modes (for $B_a>0$), with dispersion relation
 \begin{equation}
 \label{omegaLpm}
 \omega_{L,\pm}=\pm q\sqrt{\frac{B_a}{\rho_0}} + \imath \frac{q^2}{2\rho_0} \left( \eta_L- \frac{\zeta_2 \Delta\mu}{k_u}  \right)\;.
 \end{equation}
In contrast to the case of a passive gel or a gel with static bound motors, these oscillatory density waves can now become unstable when the (negative) viscosity induced by the motors overcome the internal viscous dissipation of the network, i.e., for $\zeta_2\Delta\mu\tau_{on}\geq\eta_L$.   Above the critical value of activity defined by the vanishing of the damping in Eq.~\eqref{omegaLpm}, the propagating waves become unstable and the uniform state is presumably replaced by a state that supports spontaneous oscillations.

\subsection{Transverse Modes}
The transverse equations ~(\ref{uT}) and (\ref{vT}) do not couple to motor dynamics. They
yield a cubic eigenvalue equation. There are therefore  three transverse modes in the system. Of these two are propagating shear waves, with dispersion relation for small $q$  given by
\begin{equation}
\omega({{\bf q}})=\pm q \sqrt{\frac{\mu}{\rho_g}}  - \frac{\imath q^2}{2\rho_g}\left(
\eta+\eta_s+\frac{\mu \rho_f^2}{\Gamma\rho_g} \right) +\ O(q^3)\;, \label{omegav}
\end{equation}
with $\rho_g=\rho_0+\rho_f$ the mass density of the gel. The third transverse mode is a non-hydrodynamic mode with  a finite decay rate at ${\bf q}=0$. It describes the  relative motion of the polymer network and the permeating fluid. The dispersion relation is given by
\begin{equation}
\omega({\bf q})=-\frac{\imath \Gamma\rho_g}{\rho_0\rho_f} - \frac{\imath q^2}{\rho_g}\left(  \frac{\eta\rho_0^2+\eta_s \rho_f^2}{\rho_0\rho_f} - \frac{\mu \rho_f^2}{\Gamma\rho_g}\right) +\ O(q^4)\;.
\end{equation}
Transverse fluctuations always decay and to linear order do not
destabilize the stationary homogeneous state. Finally, if $B$ and
$\mu$ are comparable, the speed of propagation of the transverse
waves given in Eq.~(\ref{omegav}) is generally much smaller than
that of the longitudinal waves given in Eq. ~(\ref{omega-pm0}),
since  $\rho_f >> \rho_0$.

\section{Overdamped Dynamics of the Polymer Network : Connection to Muscle Sarcomeres}
 In the overdamped limit of large friction $\Gamma$, the inertial term in Eq.~(\ref{ud}) is negligible and  the relaxational dynamics of the fiber density is controlled by the viscous coupling to the permeating fluid. This is the limit that is relevant to most biological systems, such as muscle sarcomeres. We show here that in this limit the on/off dynamics of bound motor yields an effective inertia that results in spontaneous oscillations even in this overdamped limit.

 The approximation of neglecting the inertial terms can be quantified as follows. The inertial term in Eq.~(\ref{ud}) can be neglected relative to the frictional damping from the fluid provided $\rho_0\omega^2<<\Gamma\omega$ or
 $\omega<<\Gamma/\rho_0\sim\eta/(\xi^2\rho_0)$, which is simply the condition of low Reynolds number for an object of typical size $\xi$ moving in a medium of kinematic viscosity $\eta/\rho_0$ at a typical speed $\sim \xi\omega$. A sarcomere of typical rest length $\xi \sim 2.5\ \mu m$~\cite{AlbertsBook07}, moves in an ambient viscous
 medium of viscosity $\eta \sim 10\ pN\ s\ \mu m^{-1}$. The mass density $\rho_0$ of a sarcomere is approximately $10^3 kg~m^{-3}$~\cite{Denoth2002}. Inertial effects can be neglected if the velocity of a sarcomere unit, typically of order $10\ \mu m\ s^{-1}$, is small compared to $ \eta/\xi \rho_0$. From the known values of sarcomere parameters, as quoted above, $\eta/\xi \rho_0 \sim 10^{-2} ms^{-1}$, which is three orders of magnitude higher than the typical velocity of a sarcomere. Hence the ignoring of the inertial forces is justified.

 A sarcomere chain can be described as a one dimensional elastic system in terms of a displacement field $u(z,t)$, with  $z$ the coordinate along the sarcomere's length.  In the overdamped limit the equation for the displacement field and the deviation of the fraction of bound motor from the steady state value  are given by
\begin{eqnarray}
&&\left(\Gamma-\eta_L\partial_z^2\right)\partial_t u=B_a\partial_{z}^2 u +\zeta_2\Delta\mu\partial_z\phi\;,\label{u1}\\
&&\partial_t\phi=- \partial_{z}\left[\left(1+\phi\right) \partial_t{u}\right] - k_{u}\phi\;.\label{motor1}
\end{eqnarray}
We note that in the overdamped limit discussed in this section our
model is formally similar to the model introduced by Murray and
Oster~\cite{MurrayOster1984} to describe the role of the
mechanochemistry of the cytogel in epithelium movements (albeit with
calcium dynamics taking the place of motor dynamics), but with one
important difference: here we consider a gel frictionally coupled to
a permeating fluid, while Refs.~\cite{MurrayOster1984,Odell1981}
consider a gel elastically coupled to a substrate. As shown below,
both models yield oscillations and traveling waves.

When linearized by approximating the convective term on the right
hand side of Eq.~\eqref{motor1} as $\sim-\partial_{z}\partial_t{u}$,
these equations are identical to those derived by
G\"{u}nther-Kruse~\cite{GuntherKruse2007} from a microscopic model
of muscle sarcomeres. Here we show that the same equations can be
obtained by a purely phenomenological approach that includes both
the dissipation due to the coupling to the permeating fluid and the
on/off motor dynamics. We also note that the bound motor fraction
can be eliminated from  the linearized equations by transforming
them  into a single differential equation for the displacement.
Solving the linearized form of Eq.~(\ref{motor1}) for $\phi$ with
$\phi(z,t=0)=0$, substituting in Eq~(\ref{u1}) and differentiating
with respect to time, we obtain a single differential equation for
the displacement $u(z,t)$, albeit second order in time, given by
\begin{equation}
\tau_{on}\left(\Gamma-\eta_L\partial_z^2\right)\partial_t^2 u +
\left[\Gamma -\eta_L\partial_z^2-\eta_a\partial_z^2\right]\partial_t
u=B_a\partial_z^2 u \label{inertia}
\end{equation}
where
\begin{equation}
\eta_a=\tau_{on}\left[B-\left(\zeta_1+\zeta_2\right)\Delta\mu\right]\;.
\end{equation}
 It is clear from Eq.~(\ref{inertia}) that the effect of motor on/off dynamics is to provide an "inertial" contribution to the dynamics of the network. On length scales large compared to $\xi_d$ we can neglect the internal dissipation  intrinsic to the network proportional to the viscosity $\eta_L$ compared to the friction $\Gamma$ with the permeating fluid. Eq.~(\ref{inertia}) then simplifies to
\begin{equation}
\tau_{on}\Gamma\partial_t^2 u + \left[\Gamma
-\eta_a\partial_z^2\right]\partial_t u=B_a\partial_z^2 u
\label{inertia2}
\end{equation}
In this limit Eq.~\eqref{inertia2} describing deformations of the
active network is formally identically to
Eq.~\eqref{deltarho-passive} for the passive gel, with
$\tau_{on}\Gamma$ playing the role of a mass density, and a
viscosity $\eta_a$ and an elastic modulus $B_a$, both renormalized
by activity. The effective viscosity and the elastic modulus can
change sign at high activities, yielding instabilities.

First we consider the hydrodynamic modes of the systems described by the linearized form of Eqs.~\eqref{u1} and \eqref{motor1} or by Eq.~\eqref{inertia}. These are given by the solutions of the  eigenvalue equation, given by
 \begin{equation}
 \omega^2(\Gamma + \eta_L q^2)+\imath \omega k_{u}\left[\Gamma+\left(\eta_L +\eta_a\right)q^2\right]  -k_{u}B_aq^2=0 \;.
 \label{eig_damped}
 \end{equation}
 The general solutions of the eigenvalue equation are
 \begin{eqnarray}
\lefteqn{ \!\!\!\!\omega=\frac{ k_u}{2(\Gamma+\eta_Lq^2)}\Big\{-\imath\left[\Gamma+\left(\eta_L +\eta_a\right)q^2\right]}\nonumber\\
 && \pm\sqrt{-\left[\Gamma+\left(\eta_L +\eta_a\right)q^2\right]^2+\frac{4B_a q^2}{k_{u}}
 (\Gamma+\eta_Lq^2)}\Big\}
 \label{geneig}
 \end{eqnarray}
For small wavevector ($q\rightarrow 0$) we obtain two modes,
 \begin{eqnarray}
 &&\omega_b=-\imath k_{u}+\imath\frac{\zeta_2\Delta\mu}{\Gamma}q^2\\
 &&\omega_L=-\imath\frac{B_a}{\Gamma}q^2
 \end{eqnarray}
 describing motor and network density relaxation, respectively. Again, the system exhibit a contractile instability when $B_a<0$, but there are no oscillatory waves in the long wavelength limit.

\begin{figure}
\begin{center}
\includegraphics[scale=0.5]{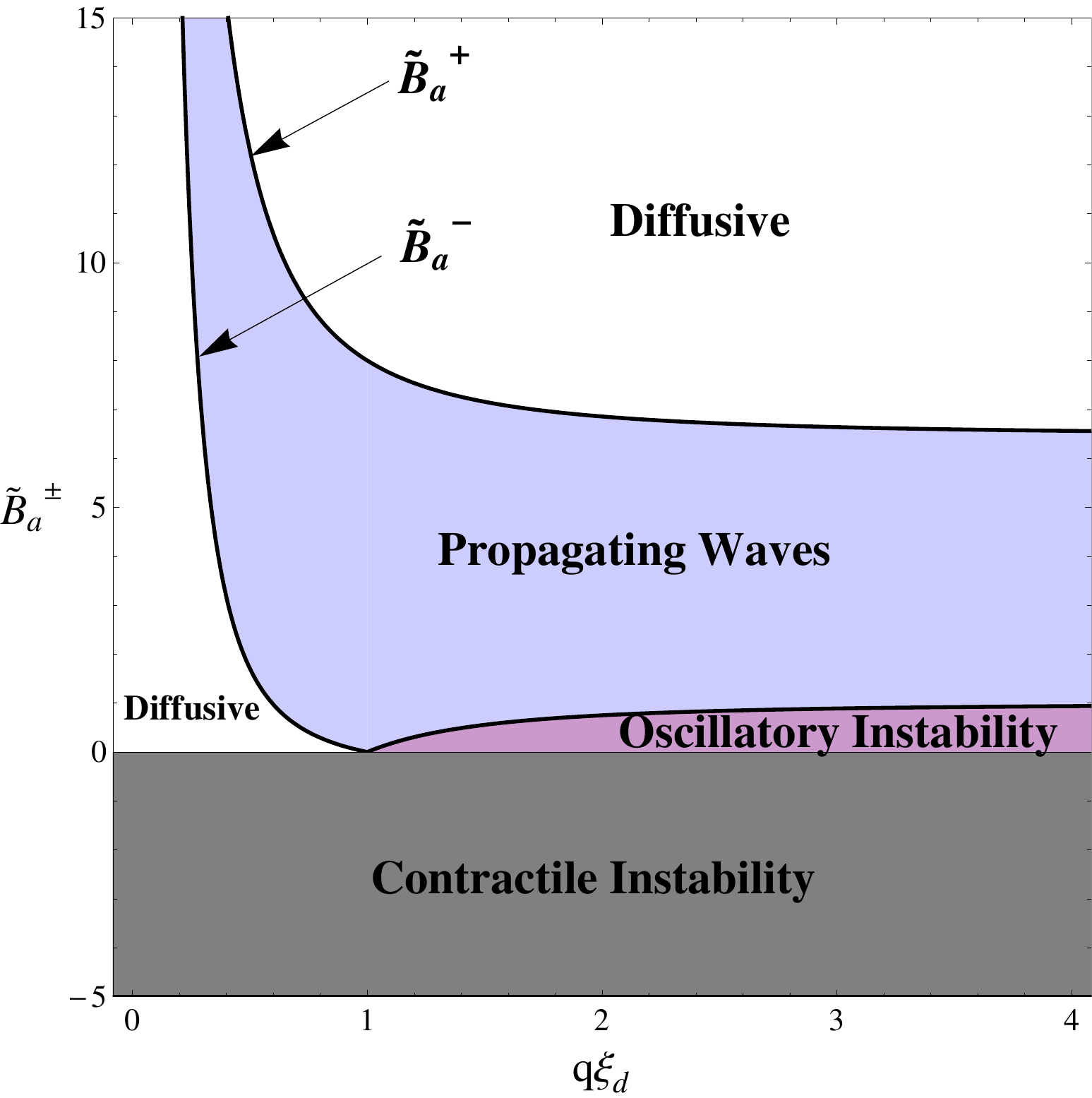}
\end{center}
\caption{ A phase diagram for the overdamped active gel. The
vertical axis is $\tilde{B}_a=B_a/(\Gamma\xi_d^2k_u)$ and the
horizontal axis is $q\xi_d$. The boundaries separating the regions of
diffusive relaxation of network density fluctuations from the region
where traveling waves exist are given by Eq.~\eqref{Epm}, plotted
here for $\tilde{\zeta}_2=2$. Below the horizontal line
$\tilde{B}_a=0$, the system exhibits a contractile instability. }
\label{fig:PDoverdamped}
\end{figure}

Propagating wave solutions exist if the argument of the square root
on the right hand side of Eq.~\eqref{geneig} is positive. The active
viscosity can be written as $\eta_a=(B_a-\zeta_2\Delta\mu)/k_u$,
hence it  depends on the renormalized elastic modulus $B_a$. If we
choose to treat $\tilde{B}_a=B_a/(\Gamma\xi_d^2k_u)$  and
$\tilde{\zeta}_2=\zeta_2\Delta\mu/(\Gamma\xi_d^2k_u)$ as independent
parameters the condition for existence of propagating waves can be
written as $\tilde{B}_{a}^-(\tilde{q})\leq
\tilde{B}_a\leq\tilde{B}_{a}^+(\tilde{q})$, with
\begin{equation}
\tilde{B}_{a}^\pm(\tilde{q})=\frac{1}{\tilde{q}^2}\left[1+\tilde{q}^2+\tilde{\zeta}_2\tilde{q}^2
\pm
2\sqrt{\tilde{q}^2(1+\tilde{q}^2+\tilde{\zeta}_2\tilde{q}^2)}\right]\;,
\label{Epm}
\end{equation}
where we assumed $\eta_L\sim\Gamma\xi_d^2$. Propagating waves then
exist in a band in the $ (\tilde{B}_a,\tilde{q})$ plane, as shown in
Fig.~\ref{fig:PDoverdamped}. The width of the band is
$\Delta\tilde{B}_a=4\sqrt{1+\tilde{\zeta}_2+1/\tilde{q}^2}$. It
vanishes at small wavevectors and goes to the constant value
$4(1+\tilde{\zeta}_2)^{1/2}$ at large wavevectors. In contrast to
the propagating density waves obtained in a damped passive gel, the
oscillatory behavior results here from motor activity and the range
of parameter where it exists grows  with the time $\tau_{on}$ that
characterizes motor dynamics. Since $\tau_{on}\sim1/\Delta\mu$ to
leading order $\tilde{\zeta}_2$ is independent of activity for small
activity. In addition, the propagating waves are unstable when the
imaginary part of the eigenvalues given by  Eq.~\eqref{geneig}  is
positive. This corresponds to
\begin{equation}
\tilde{B}_a\leq \tilde{\zeta}_2-\frac{1+\tilde{q}^2}{\tilde{q}^2}
\label{unstable-waves}
\end{equation}
and defines a region where the overdamped active gel exhibits an
oscillatory instability. We expect that when nonlinear terms are
included in the equations,  the gel will exhibit spontaneous
oscillations in this region of parameters.
The transition from diffusive to oscillatory behavior is controlled
by the interplay between $\tau_{on}$ and the characteristic time
$\tau_d\sim \xi_d^2\Gamma/B$ for the diffusive relaxation of a network
fluctuation of size $\xi_d$. If $\tau_{on}\gg\tau_d$ the on/off motor
dynamics  provides an "inertial drag" to the network that opposes
the elastic restoring forces, yielding propagating waves.
Alternatively, the result can be understood in terms of  two length
scales in the problem, $\xi_d$ and
$l_b\sim\sqrt{B/(k_u\Gamma)}$. If $\xi_d>l_b$ then density relaxation
is always diffusive in the range of wavevectors ($q\xi_d\ll 1$)
described by the present theory. If in contrast $l_b>\xi_d$ the
network supports propagating density waves in the wavevector range
$l_b^{-1}\leq q\leq \xi_d^{-1}$.

\section{Linear Response}
\subsection{Dynamic Compressional Moduli} In this section we characterize the macroscopic homogeneous viscoelastic response of the active
gel in frequency space in terms of the dynamical compressional
modulus. To describe a traditional compressional experiment, we consider a slab the
three-fluid active gel model with only longitudinal degrees of
freedom, held between two plates at $z=0$ and $z=L$ and unbounded in the other two directions.
We imagine applying a harmonic compressive strain at one end,
where $u(z=L)=u_0e^{-\imath \omega t}$, while holding the other end
 fixed, i.e., $u(z=0)=0$. In general,  both the cases
of an oscillating boundary that is permeable or impermeable to the permeating fluid are experimentally relevant.
To implement a calculation that allow to treat both cases  one needs to include a finite
compressibility so that the longitudinal elasticity equations couple to the
fluid velocity $v$. Here we limit ourselves to a permeable boundary
and impose no boundary conditions on $v$. With these boundary conditions we calculate the stress   $\sigma(z=L)$
required at the oscillating boundary and define the complex
compressional modulus $B_{expt}(\omega)$ measured in
experiments as the ratio of the
stress to the applied compressional strain, $u_0/L$.
We will see below that at low frequency we recover
the complex bulk compressional modulus, $B(\omega)=B-\imath\omega\eta_L$,
obtained assuming an affine compression over the entire sample.

First we analyze for comparison the case of the passive gel with inertia and
damping. The elastic response is  governed by the  equation
\begin{equation}
\rho_0\partial_t^2 u + \Gamma\partial_t u = B \partial_z^2 u +
\eta_L \partial_t \partial_z^2 u\;.
\end{equation}
We assume a solution of the form $u(z,t)=f(z)e^{-\imath\omega t}$,
where $f(z)=f_i e^{\lambda_i z}$, yielding  a
characteristic equation for the eigenvalues $\lambda$,
\begin{equation}
\lambda^2=-\frac{\omega^2 \rho_0 + \imath \omega \Gamma}{B(\omega)}
\end{equation}
Boundary conditions, $f(0)=0$ and $f(L)=u_0$ lead to the solution,
\begin{equation}
f(z)=u_0 \frac{\sinh{(\lambda z)}}{\sinh{(\lambda L)}}
\end{equation}
The complex dynamic compressional modulus is then given by
$B_{exp}(\omega)=\frac{L}{u_0}B(\omega)\left(\frac{df}{dz}\right)_{z=L}$ which gives
\begin{eqnarray}
B_{expt}(\omega)
=B(\omega)\lambda L \coth{(\lambda L)}
\end{eqnarray}
The eigenvalue can be written as
\begin{equation}
\lambda^2 L^2 =- \left[\frac{\omega L }{ v_s(\omega)}\right]^2 + i
\left[\frac{L}{\delta(\omega)}\right]^2
\end{equation}
 where we have defined the
frequency dependent sound speed,
$v_s(\omega)=\sqrt{B(\omega)/\rho_0}$, and the penetration depth
$\delta(\omega)=\sqrt{B(\omega)/\omega\Gamma}$ which controls the
penetration of rarefaction/compression waves of frequency $\omega$~\cite{LevineMackintosh2009}.
At low frequency, where $\vert\lambda L\vert \ll 1$, we recover
$B_{expt}(\omega) \rightarrow B(\omega)$, provided
$\omega L / v_s(\omega) \ll 1$ and $ L \ll
\delta(\omega)$. The first condition means that the frequency of applied
oscillations is small compared to the
frequency of sound wave propagation across the entire sample. When this is not satisfied
 there is an appreciable time lag
between the imposed deformation at one end of the sample and the
deformations realized at other material points across the sample, resulting in
nonuniform strain and preventing the experimental
determination of a macroscopic compressional modulus. The
second condition demands that the boundary compressional waves fully penetrated  the sample,
which is again necessary to achieve a  uniform compressional strain. For a similar discussion of
shear rheological experiments see Appendix C
of Ref.~\cite{Levine2001}.
Finally, the compressional modulus to second order in frequency
as measured in a macroscopic experiment is given by
\begin{equation}
B_{expt}(\omega)= B-\omega^2 \rho_0 L^2/3 - \imath \omega ( \eta_L + \Gamma L^2/3)
  + O(\omega^3)
\end{equation}

\begin{figure}
\begin{center}
\includegraphics[scale=0.7]{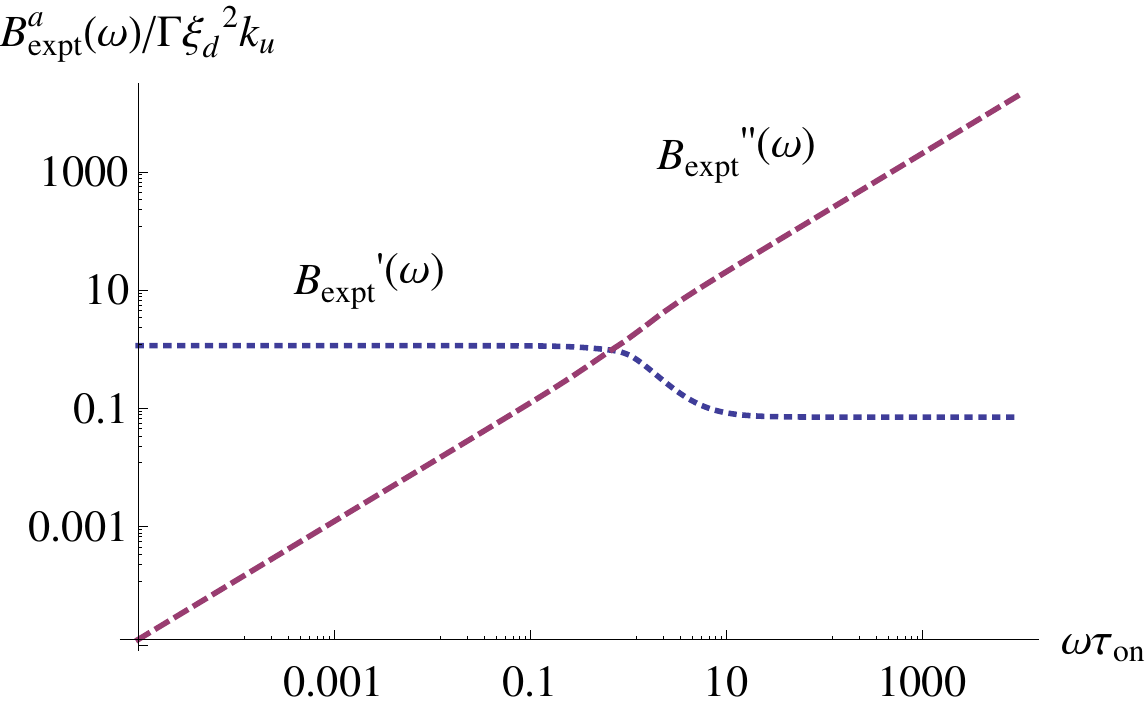}
\end{center}\caption{Storage ($B_{expt}'(\omega)$) and loss
($B_{expt}''(\omega)$) moduli,
for $\tilde{B}_a=1.15$, $\tilde{\zeta}_2=1.1$ and $\xi_d/L=0.5$.}\label{fig5}
\end{figure}

We now turn to the compressional response of an active gel.   In this case we ignore
the inertial contributions relative to the damping from the permeating fluid and look for solutions of
the linearized version of Eqs.~\eqref{u1} and \eqref{motor1} of the form
$u(z=L)=u_0e^{-\imath \omega t}$ and
$\phi(z,t)=g(z)e^{-\imath\omega t}$, with $f(z)=f_i e^{\lambda_i
z}$ and $g(z)=g_i e^{\lambda_i
z}$. The  eigenvalues are given by
%
\begin{eqnarray}
\lambda^2L^2
= -i \left[\frac{L}{\delta_a(\omega)}\right]^2\left[1 + \frac{\imath\omega \zeta_2\Delta\mu/B_a(\omega)}{-\imath\omega + k_u}\right]^{-1}\;,
\end{eqnarray}
where $\delta_a(\omega)=\sqrt{B_a(\omega)/\omega\Gamma}$
and $B_a(\omega)=B_a-i\omega\eta_L$.
Using, $-\imath\omega f'(z) = (-\imath \omega + k_u)g(z)$ and the
boundary conditions on $f(z)$ and proceeding as in the passive case, we obtain
\begin{eqnarray}
B_{expt}^a(\omega)
=\left[B_a(\omega)  + \zeta_2 \Delta\mu \frac{i\omega\tau_{on}}{1 - i\omega\tau_{on}}\right] \lambda L \coth{(\lambda L)} \;.
\end{eqnarray}
The real and imaginary parts of $B_{expt}^a(\omega)=B'_{expt}(\omega)-i B''_{expt}(\omega)$ representing the storage and loss moduli, respectively,  are shown in Fig.(\ref{fig5}) for generic values of parameters. The storage or elastic modulus has a frequency independent plateau at
frequencies lower than the motor's unbinding rate,  indicating that the system behaves
like an elastic gel in this region. The linear frequency dependence
of the loss modulus is the hallmark of a dissipative gel. At
low frequency $B'_{expt}>B''_{expt}$ and the system behaves elastically, while at
high frequency $B''_{expt}>B'_{expt}$ and the response is dominated by viscous
losses. This response is reminiscent of the Kelvin-Voigt model of viscoelasticity.
Finally, at low frequency the compressional modulus is given by
\begin{eqnarray}
B_{expt}^a(\omega)&=&B_a -
\omega^2\tau_{on}^2\zeta_2\Delta\mu
- i \omega \left(\eta_L + \frac{\Gamma L^2}{3} -
\tau_{on} \zeta_2 \Delta\mu\right)  \nonumber\\
&&+ O(\omega^3)\;,
\end{eqnarray}
whereas, at high frequencies since $\lambda \sim \sqrt{\Gamma/\eta_L}=1/\xi_d$, we obtain
\begin{equation}
B_{expt}^a(\omega) \sim \left(B_a - \zeta_2 \Delta \mu -
\imath\omega\eta_L\right) (L/\xi_d)\coth{(L/\xi_d)}\;.
\end{equation}

\subsection{Creep}
Here we study the macroscopic behavior of our active elastic
medium by considering the creep response, i.e., the time evolution
of the average strain $\varepsilon(t)=1/L\int_0^L dz\ \partial_zu$ in response  to
a homogeneous external stress, $\sigma(t)$. In particular we are interested in
characterizing the load and recovery creep of the material following
the sudden application and removal, respectively, of a constant
stress. Both responses are measured experimentally in
cells~\cite{Bartel1997,Thoumine1997,Mitrossilis2009}.

Consider a muscle fiber of length $L$ with free boundary conditions at the ends $z=0$ and $z=L$, i.e. $\partial_z u(z=0,L)=0$, and no fluctuation in motor densities being imposed at the ends. Hence one assumes normal mode expansions for $u$ and $\phi$ to be of the form $u(z,t)=\sum_{m=0}^{\infty} u_m(t) \cos{(\hat{m} z)}$, and $\phi(z,t)=\sum_{m=1}^{\infty} \phi_m(t) \sin{(\hat{m} z)}$, where $\hat{m}=m\pi/L$.

Neglecting nonlinearities, the evolution of the normal modes
$u_m(t)$ and $\phi_m(t)$ in the material in response to a small external
stress $\sigma(t)$  is governed by the equations
\begin{subequations}
\begin{gather}
(\Gamma+\eta_L\hat{m}^2) \dot{u}_m(t) + B_a \hat{m}^2 u_m(t) - \zeta_2\Delta\mu \hat{m} \phi_m(t) = f_m(t)\;,\\
\dot{\phi}_m(t)=\hat{m}\dot{u}_m(t) -k_u \phi_m(t)\;. \label{phi}
\end{gather}
\end{subequations}
With, $f_m(t)=\frac{2\sigma(t)}{L^2}\int_0^L dz \sin{(\hat{m} z)}$.

Eliminating the fluctuations $\phi_m(t)$ in the density of bound
motor, we obtain an effective equation for $u_m(t)$, given
by
\begin{equation}\label{ep1}\begin{split}
\tau_{on}(\Gamma+\eta_L\hat{m}^2) \ddot{u}_m(t) + \left[\hat{m}^2\left(\tau_{on}(B_a -
\zeta_2\Delta\mu) + \eta_L\right) + \Gamma \right] \dot{u}_m(t) &\\ + B_a \hat{m}^2 u_m(t) = \tau_{on} \dot{f}_m + f_m\;,
\end{split}\end{equation}
where $\tau_{on}=k_u^{-1}$. The decay rates of the individual modes are
\begin{equation}
\label{m-modes}
2 \gamma(m) = k_u + \frac{B_a - \zeta_2\Delta\mu}{\eta_L\left(1+\frac{L^2}{m^2\xi_d^2\pi^2}\right)}\;.
\end{equation}
Eq.~\eqref{m-modes} shows that $\gamma(m)$ is an increasing function of $m$, hence the higher modes decay at a faster rate. For simplicity we then consider only the first mode, $m=1$. Thus we approximate the averaged strain developed in the material as $\varepsilon(t) \simeq -2u_1\pi/L$. Also note that neglecting viscous coupling to the fluid $\Gamma$ amounts to considering the limit of the fastest mode $m\rightarrow \infty$.

In the limit $\tau_{on} \rightarrow 0$,
when motors are unbound at all times, Eq.~(\ref{ep1}) reduces to the
familiar Kelvin-Voigt viscoelastic equation~\cite{Wineman2000}. In
this case the creep following application of a sudden load at $t=0$,
$\sigma(t)=\sigma_0\Theta(t)$ has the familiar form
$$\varepsilon(t)=\frac{8\sigma_0/\pi}{\eta_L\pi^2+ \Gamma L^2}(1-e^{-t/(\tau_B+\Gamma L^2/\pi^2 B_a)})\;,$$ where
$\tau_B=\eta_L/B_a$ is the Kelvin-Voigt relaxation time.

\begin{figure}
\begin{center}
\includegraphics[scale=0.9]{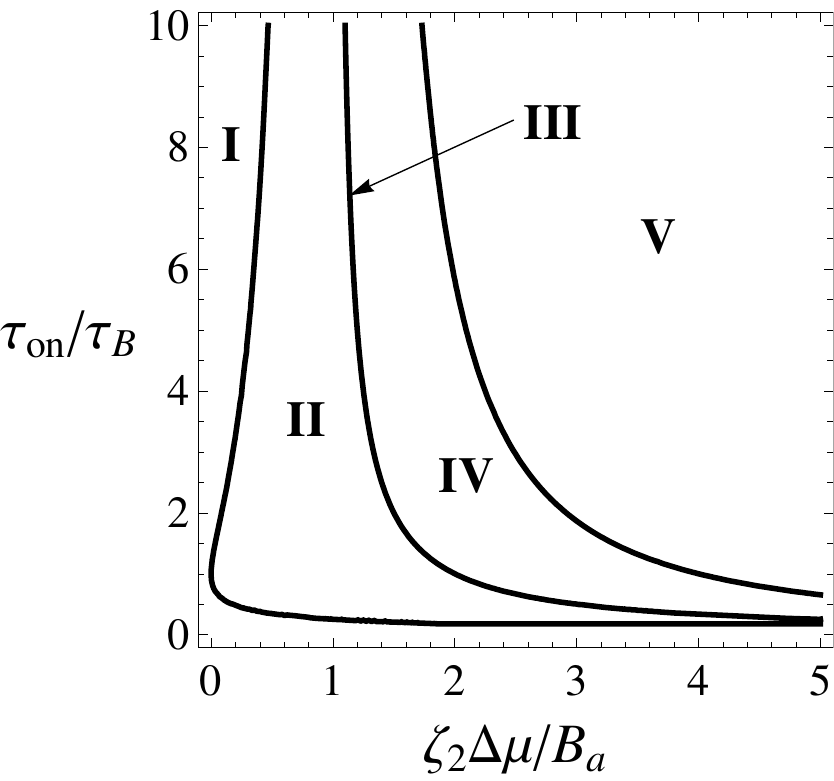}
\end{center}
\caption{A phase diagram displaying the various types of creep
response obtained for $B_a>0$. I) Stable Monotonic Decay, II) Stable
Oscillatory, III) Line of Sustained Oscillations, IV) Unstable
Oscillatory Growth, V) Unstable Monotonic Growth. }
\label{fig:creep-PD}
\end{figure}

\begin{figure}
\begin{center}
     \subfigure[ ]{
          \label{fig:stable-creep}
\includegraphics[width=0.38\textwidth]{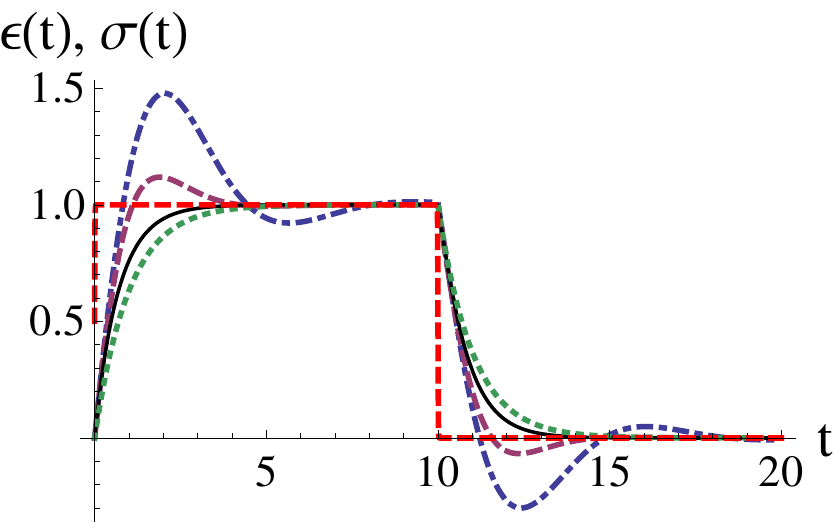}}
\subfigure[ ]{
          \label{fig:unstable-creep1}
          \includegraphics[width=.38\textwidth]{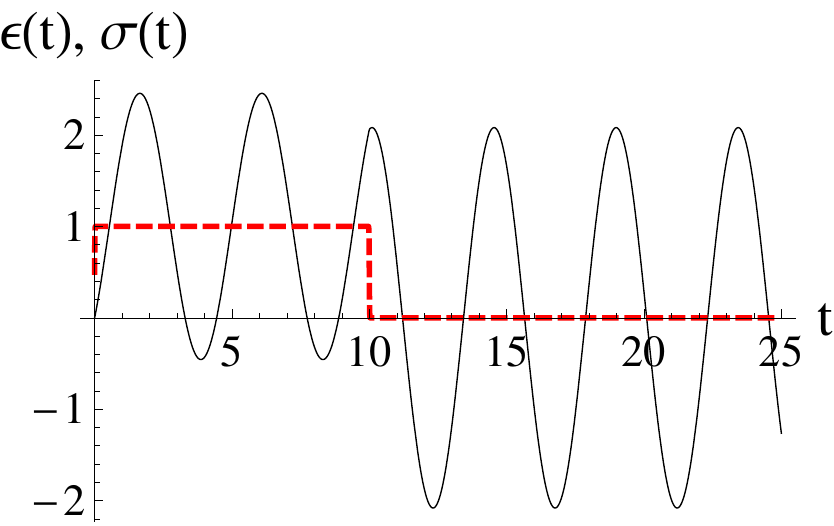}}
\end{center}
\caption{(color online) Strain $\varepsilon(t)$ in response to a
step-stress $\sigma(t)$ (dashed line, red online) with $T=10$.
Strain and stress are measured in units of $\sigma_0$ and
$\sigma_0/B_a$, respectively and time is in unites of $\tau_B$. The
various curves correspond to different values of $\tau_{on}$,
spanning the regimes described above. Top frame: $\tau_{on}=0$
(dotted line, green online), corresponding to a passive gel with
Kelvin-Voigt response; $\tau_{on} / \tau_{B}=0.2$ (solid line, black
online), corresponding to region I of monotonic stable response;
$\tau_{on} / \tau_{B}=0.5$ (dashed line, purple online), and
$\tau_{on} / \tau_{B}=1$ (dashed-dotted line, blue online),
corresponding to region II of oscillatory stable response. All
curves are for $\zeta_2/B_a=1$. Bottom frame:
$\tau_{on}/\tau_{B}=0.5$ and  $\zeta_2/B_a=3$, corresponding to
region III of sustained oscillations. } \label{fig:creep}
\end{figure}

For finite values of $\tau_{on}$, the creep response is controlled
by the interplay of the two times scales $\tau_B$ and $\tau_{on}$.
We assume $B_a >0$, corresponding to weak activity. When $B_a<0$ the
system exhibits  a contractile instability and the strain becomes
arbitrarily large at long times for any applied $\sigma(t)$.
%
%
The evolution of the strain in response to an applied
stress is then controlled by the two eigenvalues  of Eq.~\eqref{ep1} for $m=1$, given by
\begin{equation}
\lambda_{\pm}=-\gamma\pm\sqrt{\gamma^2-\frac{\tau_B}{J(L)\tau_{on}}}\;,
\end{equation}
where time is measured in units of $\tau_B$, and
$$2\gamma=\left(1 - \frac{\zeta_2\Delta\mu}{B_a} +
\frac{\tau_B}{\tau_{on}}+\frac{\Gamma L^2}{\pi^2B_a \tau_{on}}\right)/J(L),$$
$$J(L)=\left[1+\left(\frac{L}{\pi\xi_d}\right)^2\right].$$
The linear creep response of the active gel can then be classified as follows:
\begin{description}
\item {I.} $\gamma > 0$ , $\gamma^2 > \frac{\tau_B}{J(L)\tau_{on}}$ : stable monotonic behavior
\item {II.} $\gamma > 0$ , $\gamma^2 < \frac{\tau_B}{J(L)\tau_{on}}$ : stable oscillatory behavior
\item{ III.} $\gamma=0$ : sustained oscillations
\item{ IV.} $\gamma < 0$ , $\gamma^2 > \frac{\tau_B}{J(L)\tau_{on}}$ : unstable oscillatory growth
\item{ V.} $\gamma < 0$ , $\gamma^2 < \frac{\tau_B}{J(L)\tau_{on}}$ : unstable monotonic growth
\end{description}
The behavior is summarized in the phase diagram of Fig.~\ref{fig:creep-PD}
displaying the various regions in the $(\tau_{on},\zeta_2)$ plane for fixed $J(L)$.
We note that when $\tau_{on}/\tau_{B}$ is increased for fixed
$\zeta_2\Delta\mu/B_a\geq 1$, the material eventually becomes
unstable to stretching.
Figures ~\ref{fig:stable-creep}  and \ref{fig:unstable-creep1} show
the time evolution of the strain in response to a step stress of
height $\sigma_0$ and duration $T$,
$\sigma(t)=\sigma_0\left[\Theta(t)-\Theta(t-T)\right]$, with initial
condition $\varepsilon(0)=0$. The response in region I of stable
monotonic decay is similar to conventional Kelvin-Voigt response. In
region II of stable oscillatory decay  the interplay of the two time
scales $\tau_{on}$ and $\tau_B$ yields the possibility of a strain
overshoot.  For finite $\tau_{on}$ we also need to specify an
additional initial condition determined by the initial distribution
of bound motors, since
$\dot\varepsilon(0)=\dot\phi(0)-\phi(0)/\tau_{on}$. Fig.~
\ref{fig:unstable-creep1} displays the response in region III of
sustained oscillations.

\section{Discussion and conclusions}

We have presented a generic continuum theory of active gels, modeled
as a viscoelastic solid with bound motor proteins that induce active
stresses in the medium. In the limit where the inertia of the network is
neglected and the equations are specialized to one dimension, the
model is equivalent to that proposed by G\"unther and Kruse~\cite{GuntherKruse2007} by
coarse-graining of a specific mechanical model of coupled muscle
sarcomeres. For large values of the motor activity as
measured by the rate of ATP consumption, $\Delta\mu$, the
contractile action of bound motors yields a diffusive (contractile)
instability of the gel. This result has been obtained earlier in
models of muscle sarcomeres~\cite{GuntherKruse2007} and actin bundles~\cite{Schaller2008}. Here we  show
that it is a generic property of active elastic media. For smaller
values of motor activity the interplay of solid elasticity and the
binding/unbinding dynamics of the motor proteins yields propagating
waves and eventually oscillatory instabilities in the linear theory.
Both stable and unstable oscillatory modes are obtained even in the
case of an overdamped gel, as relevant to muscle fibers. We show
that the finite time scale of motor on/off dynamics yields an
effective "inertial" contribution to the dynamics of the elastic
medium controlled by the time $\tau_{on}$ that motors spend bound to
filaments (see Eq.~\eqref{inertia}). One of the new results of the
paper is the phase diagram displayed in Fig.~\ref{fig:creep-PD} for
the macroscopic response of the system to external stresses. In the
linear model sustained oscillations are only obtained for special
parameter values corresponding to a line in the
$(\tau_{on},\zeta_2)$ phase diagram. It is expected that
nonlinearities neglected in the present work will  have a
stabilizing effect and  replace the unstable oscillatory response
with stable self-sustained oscillations. The model considered is
relevant for the description of motor-induced spontaneous
oscillations in muscle sarcomeres and other active elastic media,
and may provide a useful framework for the understanding of
lamellipodium crawling.

We plan to extend this work in various ways.
First, an analysis of the effect on nonlinearities is needed. Two
classes of nonlinear terms are important in our model of an active
gel. The first is provided by nonlinear convective terms in the
equation describing the dynamics of bound motors, as shown in
Eq.~\eqref{motor1}, and also including dependence of the unbinding rate $k_u$ on the elastic strain $\partial_z u$ developed in the gel. These are the simplest continuum manifestation
of the highly nonlinear load dependence of the microscopic motor
unbinding rate, which in turn plays an important role in controlling
the motor-induced negative friction induced by the cooperative
action of motor proteins on biological systems  elastically coupled
to their environment~\cite{JulicherProst1995,Grill2005,Placais2009}.
A second class of nonlinearities arise from higher order terms in
the expansion of the active parameter $\zeta$ given in
Eq.~\eqref{actsigma}. A preliminary estimate of the effect of these
terms suggest that they stabilize the oscillatory growing modes
and yield stable sustained oscillations.

In the liquid state of an active system the polarity of actin
filaments plays an important role. The coupling of polarity and flow
has been shown to yield spontaneous flow~\cite{Voituriez06}, banded
states of inhomogeneous concentration, and oscillatory
states~\cite{GiomiMarchettiLiverpool2008}. It is similarly expected
that the coupling of polarity and elasticity will yield new
phenomena in active solids, including spontaneous deformations and
oscillations. To incorporate the effect of polarity we have begun to
consider the properties of an active polar elastomer, where the
orientational order can be induced either by elongated passive
crosslinkers~\cite{Dalhaimer2007} or by the myosin minifilaments
themselves. In addition, the latter exert active force dipoles on
the medium that induce  active stresses coupled to the orientational
order. A detailed discussion and analysis of such active elastomers
is left for a future publication.

\vspace{0.3in}

We acknowledge support from  the NSF on grants DMR-075105 and
DMR-0806511. We thank Aparna Baskaran, Alex Levine, Tannie Liverpool and Kristian
M\"uller-Nedebock  for illuminating discussions. SB also thanks the
University of Stellenbosch for hospitality during the completion of
part of this work.

\end{document}